Title: On a simple derivation of the very low damping escape rate for classical spins by modifying the method of Kramers

Authors: Declan Byrne, William T. Coffey, Yuri P. Kalmykov, Serguey V. Titov

Published in: Physica A: Statistical Mechanics and its Applications

Journal reference: Physica A, volume 527, 121195

Date of Publication: 1 August 2019

DOI: 10.1016/j.physa.2019.121195

Publisher: Elsevier

# On a simple derivation of the very low damping escape rate for classical spins by modifying the method of Kramers

D. J. Byrne*[1,2], W. T. Coffey[1], Yu. P. Kalmykov[3], and S. V. Titov[1,4]

[1]*Department of Electronic and Electrical Engineering, Trinity College, Dublin 2, Ireland*
[2]*Electricity Supply Board, Dublin 3, Ireland*
[3]*Laboratoire de Mathématiques et Physique, Université de Perpignan Via Domitia, 54, Avenue Paul Alduy, F-66860 Perpignan, France*
[4]*Kotel'nikov Institute of Radio Engineering and Electronics of the Russian Academy of Sciences, Vvedenskii Square 1, Fryazino, Moscow Region, 141190, Russian Federation*

*Corresponding author. E-mail address: byrned12@tcd.ie

**Abstract**

The original perturbative Kramers' method (starting from the phase space coordinates) [H.A. Kramers, *Physica* **7**, 384 (1940)] of determining the energy-controlled-diffusion equation for Newtonian particles with *separable* and *additive* Hamiltonians is generalized to yield the energy-controlled diffusion equation and thus the very low damping (VLD) escape rate including spin-transfer torque for classical giant magnetic spins with two degrees of freedom. These have dynamics governed by the magnetic Langevin and Fokker-Planck equations and thus are generally based on *non-separable* and *non-additive* Hamiltonians. The derivation of the VLD escape rate directly from the (magnetic) Fokker-Planck equation for the surface distribution of magnetization orientations in the configuration space of the polar and azimuthal angles $(\vartheta, \varphi)$ is much simpler than those previously used.

# 1. Introduction

The rate of escape of particles over potential barriers due to the shuttling action of the Brownian motion arising from their heat bath constitutes one of the famous problems of physics and chemistry. This was effectively solved by Kramers in 1940 [1] for assemblies of Newtonian particles moving in a one dimensional extension $q$, acted upon by an external conservative force $K(q) = -\partial_q V(q)$, so that they are characterized by *separable* and *additive* Hamiltonians, in the limiting cases of (a) very weak and (b) intermediate to high bath coupling. These we separately identify as very low damping (VLD) (energy-controlled-diffusion) and intermediate to high damping (IHD). The IHD regime encompasses the escape rate limits of very high damping (VHD) and intermediate damping (ID), the latter coinciding with classical transition state theory (TST) which forms the upper bound of the escape rate [2]. In both the VHD and VLD regimes, Kramers reduced the calculation of the escape rate to solving one dimensional diffusion equations. In the VLD regime, where inertial effects dominate, the diffusion equation is in energy space, while in VHD, where these effects are ignored, the diffusion equation is in configuration space and is commonly known as the Smoluchowski equation [2]. Since noise activated escape over a barrier is an exponentially slow process, it follows from the quasi-stationary solutions of both these equations that in VLD the escape rate is directly proportional to the damping coefficient while in VHD the escape rate is inversely proportional to it. The different coupling behaviour in the two limiting damping regimes then poses the Kramers turnover problem in the crossover region, where neither VLD nor IHD formulas are valid [2]. This problem was first solved many years later by Mel'nikov [3] and Meshkov and Mel'nikov [4]. They calculated the escape rate in the so-called low damping regime thereby including both the VLD and ID regimes and then extended the results in heuristic fashion to the entire IHD regime so providing a solution for all values of the damping. Later Grabert and Grabert *et al.* [5] presented a complete solution of the Kramers turnover problem and have shown that the Mel'nikov and Meshkov turnover formula can be obtained without *ad hoc* interpolation between the VLD and VHD regimes.

Versions of the Kramers escape rate theory are still being employed in innovative scientific research. For example, this theory as applied to Josephson junctions [6] has recently been used in the design of single photon detectors [7]. The current-voltage characteristics of Josephson junctions can be modelled as a Brownian particle in a tilted periodic potential, where the degree of the tilt is proportional to the bias current and the escape from a potential well corresponds to the creation of a non-zero voltage across the junction. In Ref. [7] the mean bias current required to create this voltage is calculated via the Kramers escape rate when the temperature is high enough to insure thermally activated dynamics. However, for lower temperatures (below the crossover temperature) the dynamics will be dominated by microscopic quantum tunnelling. Furthermore, this work has been expanded to consider methods of detecting the hypothetical axions and axion-like particles [8] which have been postulated to be a component of the Dark Matter of the Universe [9].

Much effort has been expended in solving the Kramers escape rate problem for classical giant spin modelling, e.g., magnetization reversal in single domain ferromagnetic nanoparticles. Here escape (i.e., reversal of the direction of precession of the moving axis of the magnetization) takes place over internal magnetocrystalline anisotropy-Zeeman energy barriers. The inverse escape rate then yields the reversal time of the magnetization of such particles [10,11]. Here the VLD damping regime is of the most practical interest for the magnetization reversal because typical dissipation parameter $\alpha$ values lie in the range $10^{-3} < \alpha < 10^{-1}$), i.e., in the one considered here [10,12]. In general, however, the spin escape rate problem differs markedly from that for Newtonian particles, rigid rotators or associated phenomena such as the current-voltage characteristics of Josephson junctions as mentioned previously in several fundamental aspects. As two degrees of freedom (namely, the polar and azimuthal angles $\vartheta, \varphi$ describing the magnetization orientation in configuration space) are involved, the spin Hamiltonian, unlike that of particles, is no longer *separable* and *additive* since the governing magnetic Langevin equation is essentially a form of the Larmor equation so that inertial effects play no role here [11,12]. Nevertheless, the role of inertia in the mechanical system is essentially mimicked in the magnetic system for noncircularly symmetric free energy potentials by the gyromagnetic term, which causes coupling or entanglement of longitudinal relaxation and transverse resonance modes. Moreover, for circular symmetry, where only one degree of freedom, $\vartheta$, is needed, since the longitudinal and transverse components of the magnetization decouple, the magnetic Fokker-Planck equation is *exact* unlike the approximate Smoluchowski equation for Newtonian particles or rigid rotators in configuration space.

We recall that the VLD spin escape rate both with and without STT has already been determined [12,13] either via involved vector manipulation [14] or else via long and complicated calculations with the energy and phase as variables in the magnetic Langevin equation involving multiplicative noise [13,15,16,17]. Both methods ultimately lead to the same energy-controlled-diffusion equation for giant classical spins with quasi-stationary solutions yielding the VLD rate. In passing, the VLD escape rate for spins (excluding STT) was first obtained by Klik and Gunther [18] using the uniform asymptotic expansion of the mean first passage time method of Matkowsky *et al.* [19]. However, this procedure also involves tedious calculations based on boundary layer theory. Here we indicate how the energy-controlled-diffusion equation for spins and thus the VLD escape rate may be obtained in far more straightforward fashion from Brown's Fokker-Planck [11,12] equation for the surface distribution of magnetization orientations in configuration space $(\vartheta, \varphi)$ by suitably adapting the VLD calculation of Kramers for particles originally based on the representation phase coordinates $(q, p)$ of the position $q$ and momentum $p$.

**2. Brown's Fokker-Planck Equation including STT**

To write down the relevant magnetic Fokker-Planck equation in terms of $(\vartheta,\varphi)$, commonly known as Brown's Fokker-Planck equation [11], we first consider the magnetic Langevin equation which may be briefly described as follows. The magnetization **M** of a ferromagnetic nanoparticle of volume $v$ precesses (without damping and thermal agitation) about an effective magnetic field **H** comprising the anisotropy and external applied fields according to the gyromagnetic equation [10]

$$\dot{\mathbf{M}}_{pr} = -\gamma(\mathbf{M}\times\mathbf{H}), \tag{1}$$

where $\gamma$ is the gyromagnetic type constant and since the effective field is the gradient of a scalar free energy density potential $V$

$$\mathbf{H} = -\frac{1}{\mu_0 M_S}\frac{\partial V}{\partial \mathbf{u}}, \quad \mathbf{u} = \frac{\mathbf{M}}{M_S}, \tag{2}$$

where $M_S$ is the saturation magnetization and $\mu_0 = 4\pi\cdot 10^{-7}\,\mathrm{JA^{-2}m^{-1}}$ in SI units. Equation (1) merely represents the kinematic equation,

$$\dot{\mathbf{u}}_{pr} = \mathbf{\Omega}\times\mathbf{u}, \tag{3}$$

with angular velocity

$$\mathbf{\Omega}(t) = -\frac{\gamma}{\mu_0 M_S}\frac{\partial V}{\partial \mathbf{u}}, \tag{4}$$

and with damping included [10,12] we have

$$\dot{\mathbf{u}} = -\gamma(\mathbf{u}\times\mathbf{H}) + \alpha(\mathbf{u}\times\dot{\mathbf{u}}), \tag{5}$$

where $\alpha$ is the dimensionless damping factor. Moreover, including spin-transfer torque (STT), we have [20]

$$\dot{\mathbf{u}} = -\gamma(\mathbf{u}\times(\mathbf{H}+\dot{\mathbf{u}}_{ST})) + \alpha(\mathbf{u}\times\dot{\mathbf{u}}), \tag{6}$$

where $\dot{\mathbf{u}}_{ST} = -\frac{1}{\mu_0 M_S}\left(\mathbf{u}\times\frac{\partial \Phi}{\partial \mathbf{u}}\right)$ and $\Phi$ is the nonconservative STT potential given by [20,21]

$$\Phi(\mathbf{u}) = J\frac{kT}{vc_P}\ln[1+c_P(\mathbf{u}\cdot\mathbf{e}_P)] \tag{7}$$

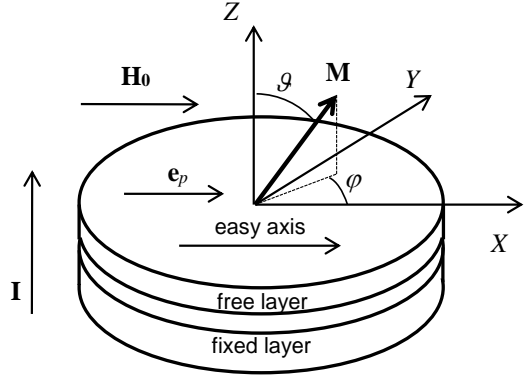

FIG. 1. Geometry of the problem. A STT device consists of two ferromagnetic strata labeled the free and fixed layers, respectively, and a normal conducting spacer all sandwiched on a pillar between two ohmic contacts. **I** is the STT current and **H**$_0$ is the external magnetic field.

and $kT$ is the thermal energy. For the typical nanopillar model (see Fig. 1) often used for spintronics the unit vector $\mathbf{e}_P$ identifies the magnetization direction in the fixed layer, $J = b_P \hbar I / (|e|kT)$ is the dimensionless STT parameter, $I$ is the current which is taken as positive if electrons flow from the free into the fixed layer, $e$ is the electronic charge, $\hbar$ is Planck's reduced constant, and the coefficients $b_P$ and $c_P$ are model dependent and are determined by the spin-polarization factor $P$ ($0 < P < 1$) [20]. For a qualitative description of STT effects, $\Phi$ may be written (by expanding the logarithm) in the following simplified form [15,16,22]:

$$\Phi(\mathbf{M}) = J \frac{kT}{vM_S} (\mathbf{e}_P \cdot \mathbf{M}). \tag{8}$$

Finally, with noise included we have the Langevin equation [11,12]

$$\dot{\mathbf{u}} = \frac{h'}{\alpha}\left(\mathbf{u}\times\left(\frac{\partial}{\partial \mathbf{u}}(V-\alpha\Phi) - \mu_0 M_S \mathbf{h}\right)\right) + h'\mathbf{u}\times\left(\mathbf{u}\times\left(\frac{\partial}{\partial \mathbf{u}}(V+\alpha^{-1}\Phi) - \mu_0 M_S \mathbf{h}\right)\right), \tag{9}$$

where $h' = \dfrac{\gamma}{(\alpha+\alpha^{-1})\mu_0 M_S}$ as defined by Brown [11] and **h** is a random magnetic field with Gaussian white noise properties

$$\langle h_i(t) \rangle = 0, \quad \langle h_i(t_2) h_j(t_2) \rangle = \frac{2kT\alpha}{v\gamma\mu_0 M_S} \delta_{ij}\delta(t_1-t_2) \tag{10}$$

where the indexes $i,j = 1,2,3$ in Kronecker's delta $\delta_{ij}$ and $h_i$ correspond to the Cartesian axes $X$, $Y$, $Z$ of the laboratory coordinate system $OXYZ$, and $\delta(t)$ is the Dirac $\delta$ function. The angular brackets mean the statistical average over an ensemble of moments which all have at time $t$ the same sharp value of the magnetization **M**. The Langevin equation (9) ultimately leads to Brown's Fokker-Planck equation for the surface probability density $W(\vartheta,\varphi,t)$ of magnetization orientations on the unit sphere incorporating STT effects [12,23],

$$\frac{\partial W}{\partial t} = \frac{kTh'}{v}\Delta W + \frac{h'}{\sin\vartheta}\left\{\frac{\partial}{\partial\vartheta}\left[\left(\sin\vartheta\frac{\partial(V+\alpha^{-1}\Phi)}{\partial\vartheta} + \frac{\partial(\alpha^{-1}V-\Phi)}{\partial\varphi}\right)W\right] \right.$$
$$\left. + \frac{\partial}{\partial\varphi}\left[\left(\frac{1}{\sin\vartheta}\frac{\partial(V+\alpha^{-1}\Phi)}{\partial\varphi} - \frac{\partial(\alpha^{-1}V-\Phi)}{\partial\vartheta}\right)W\right]\right\}. \quad (11)$$

For VLD, we are only interested in very small damping constants $\alpha$ such that we can ignore terms $o(\alpha^2)$. Furthermore, we suppose that the spin polarization current is also *small* so that all terms $J\alpha, J^2$ may also be neglected; meaning that we will restrict ourselves to a spin that would perform a *pure* precessional motion at constant energy in a well of the (for our purposes bistable) potential $V(\vartheta,\varphi)$ if no Brownian or STT torques existed. Paraphrasing Kramers [1], by small damping $\alpha$ is meant that the latter torques cause only a *very small variation of the energy during the time of one precession*. The effects of the Brownian motion and STT will therefore in their main aspect consist in the gradual change of the distribution of the axes of precession of the ensemble of spins over the different energy-values. For our purposes, on introducing the normalized free energy and STT potential,

$$E = vV/(kT) \text{ and } P = v\Phi/(kT), \quad (12)$$

the evolution Eq. (11) can be rewritten for very small $\alpha$ and $J$ (so that we ignore terms of the order of $J\alpha, J^2$, etc.) as

$$\frac{DW}{Dt} = \frac{\partial W}{\partial t} + \frac{D_L W}{Dt} = \text{St}(W), \quad (13)$$

where the left-hand side represents the nonzero hydrodynamical (convective) derivative in the Liouville equation for the surface probability density of orientations, namely

$$\frac{DW}{Dt} = \frac{\partial W}{\partial t} - \frac{1}{\sin\vartheta} \frac{\gamma kT}{v\mu_0 M_S} \left[ \frac{\partial W}{\partial \vartheta} \frac{\partial E}{\partial \varphi} - \frac{\partial W}{\partial \varphi} \frac{\partial E}{\partial \vartheta} \right] \quad (14)$$

while the right-hand side represents the effect of the dissipation and the external work done by the STT viz.,

$$\mathrm{St}(W) = \alpha \frac{\gamma kT}{v\mu_0 M_S} \left\{ \frac{1}{\sin\vartheta} \frac{\partial}{\partial \vartheta} \left( \sin\vartheta \frac{\partial W}{\partial \vartheta} \right) + \frac{1}{\sin^2\vartheta} \frac{\partial^2 W}{\partial \varphi^2} \right. \\ \left. + \frac{1}{\sin\vartheta} \frac{\partial}{\partial \vartheta} \left( W \sin\vartheta \frac{\partial}{\partial \vartheta}(E + \alpha^{-1}P) \right) + \frac{1}{\sin^2\vartheta} \frac{\partial}{\partial \varphi} \left( W \frac{\partial}{\partial \varphi}(E + \alpha^{-1}P) \right) \right\}. \quad (15)$$

### 3. Energy-controlled-diffusion equation with STT

By analogy with Kramers' derivation of the energy-controlled diffusion equation for point particles in the VLD limit [1], one may parameterize the instantaneous magnetization direction of a macrospin by the slow dimensionless energy variable $E$ and the fast precessional variable $\phi$ running uniformly along a closed Stoner-Wohlfarth orbit of energy $E$ implying that $d\phi = f_E dt$ [13], where $f_E$ is the precession frequency of precession in the potential well at a given energy $E$ [13,24]. The phase $\phi$ is the generalized coordinate conjugate to the magnetic action $S(E)$, i.e., the area inside a *closed* region of constant energy $E$ [13,24]. The magnetic action can be written in dimensionless form as [13]

$$S(E) = \frac{v\mu_0}{\gamma M_S kT} \oint_E \dot{\mathbf{M}} \cdot d\mathbf{M}. \quad (16)$$

We now introduce a new distribution function in energy and phase variables, $W(E,\phi,t)$. Mindful of the fact that both of the functions $W(\vartheta,\varphi,t)$ and $W(E,\phi,t)$ must yield the same average value for any arbitrary function $F$, namely

$$\langle F \rangle(t) = \int_0^{2\pi} \int_0^{\pi} F(\vartheta,\varphi) W(\vartheta,\varphi,t) \sin\vartheta \, d\vartheta \, d\varphi = \int_0^1 \int_{E_{\min}}^{E_{\max}} F(E,\phi) W(E,\phi,t) \, dE \, d\phi \quad (17)$$

we have

$$W(\vartheta,\varphi,t) \leftrightarrow \frac{v\mu_0 M_S}{\gamma kT} f_E W(E,\phi,t), \qquad (18)$$

where we have used the Jacobian of the transformation, namely,

$$\det\begin{pmatrix} \frac{\partial E}{\partial \vartheta} & \frac{\partial E}{\partial \varphi} \\ \frac{\partial \phi}{\partial \vartheta} & \frac{\partial \phi}{\partial \varphi} \end{pmatrix} = \sin\vartheta\left(\frac{1}{\sin\vartheta}\frac{\partial E}{\partial \vartheta}\frac{\partial \phi}{\partial \varphi} - \frac{1}{\sin\vartheta}\frac{\partial E}{\partial \varphi}\frac{\partial \phi}{\partial \vartheta}\right) \qquad (19)$$

$$= \frac{v\mu_0 M_S}{\gamma kT}\sin\vartheta\left(\dot\varphi\frac{\partial \phi}{\partial \varphi} + \dot\vartheta\frac{\partial \phi}{\partial \vartheta}\right) = \frac{v\mu_0 M_S}{\gamma kT}\dot\phi\sin\vartheta = \frac{v\mu_0 M_S}{\gamma kT}f_E\sin\vartheta.$$

The Fokker–Planck equation for $W(E,\phi,t)$ is still difficult to treat in two new state variables $E$ and $\phi$. However, since in the VLD regime, the energy $E$ diffuses very slowly over time, i.e., is almost conserved, while in contrast the phase $\phi$ varies rapidly (so that the probability density function $W(E,\phi,t)$ nearly equilibrates in $\phi$ and slowly evolves in $E$), the dependence of $W(E,\phi,t)$ on the fast variable $\phi$ may be eliminated by exploiting the periodicity of $W(E,\phi,t)$ in $\phi$ along a precessional (Stoner-Wohlfarth) orbit with a period $P_E = 1/f_E$. This is accomplished by averaging $W(E,\phi,t)$ along a *closed* trajectory of the energy over $\phi$, namely

$$W(E,t) = \overline{W(E,\phi,t)} = \int_0^1 W(E,\phi,t)d\phi \qquad (20)$$

because the phase $\phi$ changes by unity on the periodic precession along the Stoner-Wohlfarth orbit. As in the Kramers treatment for particles, denoting by $WdS$ the fraction of the ensemble enclosed by the elementary region $dS$, we may find an energy-controlled-diffusion equation for the *slightly* perturbed time dependent component of the energy distribution $W(E,t)$ by averaging Eq. (13) (written for small $\alpha$ and small STT) over $dS$ yielding

$$\overline{\frac{\partial W}{\partial t}} + \overline{\frac{D_L W}{Dt}} = \overline{\mathrm{St}(W)} \qquad (21)$$

where the overbar means averaging over the fast variable $\phi$.

First, we will justify setting the average of the Liouville term as zero in accordance with the original approach of Kramers for particles with additive Hamiltonians $H(q,p) = p^2/2m + V(q)$ placed in a bath at temperature $T$. The Liouville term averaged over the band $dS$ is now

$$\overline{\frac{D_L W}{Dt}} = \frac{\gamma kT}{v\mu_0 M_S} \overline{\frac{1}{\sin\vartheta}\left[\frac{\partial W}{\partial \varphi}\frac{\partial E}{\partial \vartheta} - \frac{\partial W}{\partial \vartheta}\frac{\partial E}{\partial \varphi}\right]}. \tag{22}$$

We transform the Liouville term to energy and phase variables, viz., $\{\vartheta,\varphi\} \to \{E,\phi\}$ so that

$$\begin{aligned}\overline{\frac{D_L W}{Dt}} &= \frac{\gamma kT}{v\mu_0 M_S}\overline{\frac{1}{\sin\vartheta}\left\{\frac{\partial W}{\partial E}\left[\frac{\partial E}{\partial \varphi}\frac{\partial E}{\partial \vartheta} - \frac{\partial E}{\partial \vartheta}\frac{\partial E}{\partial \varphi}\right] + \frac{\partial W}{\partial \phi}\left[\frac{\partial \phi}{\partial \varphi}\frac{\partial E}{\partial \vartheta} - \frac{\partial \phi}{\partial \vartheta}\frac{\partial E}{\partial \varphi}\right]\right\}} \\ &= -\overline{\frac{\partial W}{\partial \phi}\left[\frac{\partial \phi}{\partial \varphi}\dot\varphi + \frac{\partial \phi}{\partial \vartheta}\dot\vartheta\right]} = -\overline{\dot\phi \frac{\partial W}{\partial \phi}} = -f_E \int_0^1 \frac{\partial W}{\partial \phi}d\phi = 0,\end{aligned} \tag{23}$$

because $W$ is a periodic function in $\phi$, viz., $W(E,\phi+1,t) = W(E,\phi,t)$. Therefore, we have shown that the contribution of the Liouville term to the perturbed density disappears in the VLD and small STT limit. Next, on changing the variables in the term St($W$) in Eq. (13) (see Appendix A and Eq. (18)) and averaging it over the fast phase variable $\phi$, we obtain

$$\overline{\frac{\text{St}(f_E W)}{f_E}} = \alpha \frac{v\mu_0 M_S}{\gamma kT}\frac{\partial}{\partial E}\overline{\left(\frac{|\dot{\mathbf{u}}_{pr}|^2}{f_E}\left(\frac{\partial}{\partial E}(f_E W) + f_E W + \alpha^{-1} f_E W \frac{\partial P}{\partial E}\right)\right)}. \tag{24}$$

We may *factorize* the averages in Eq. (24) because the perturbations in the distribution $W$ is, by hypothesis, implicitly of order $\alpha$. Recall that in all instances, we are undertaking a perturbation to first order in $\alpha$ about a steady precessional (Stoner-Wohlfarth) orbit of energy $E$. Hence, Eqs.(21), (23), and (24) yield the energy-controlled-diffusion equation for the distribution function $W(E,t) = \overline{W(E,\phi,t)}$, viz.,

$$\frac{\partial W}{\partial t} = \frac{v\mu_0 M_S}{\gamma kT}\frac{\partial}{\partial E}\left(\frac{\alpha}{f_E}\overline{|\dot{\mathbf{u}}_{pr}|^2}\left(\frac{\partial}{\partial E}(f_E W) + f_E W\right) + \overline{|\dot{\mathbf{u}}_{pr}|^2 \frac{\partial P}{\partial E}}W\right). \tag{25}$$

Equation (25) can also be rewritten in terms of the dimensionless action $S_E$, Eq. (16), and the dimensionless work $V_E$ done by the STT along an unperturbed Stoner-Wohlfarth orbit as [13,14,16,24]

$$\frac{\partial W}{\partial t} = \alpha \frac{\partial}{\partial E}\left(S_E\left(\frac{\partial f_E W}{\partial E} + f_E W\right)\right) - \frac{\partial}{\partial E}(V_E f_E W), \tag{26}$$

where

$$S_E = \frac{v\mu_0 M_S}{\gamma kT f_E}\overline{|\dot{\mathbf{u}}_{\text{pr}}|^2} = \frac{v\mu_0 M_S}{\gamma kT f_E}\int_0^1 |\dot{\mathbf{u}}_{\text{pr}}|^2 d\phi = \frac{v\mu_0 M_S}{\gamma kT}\int_0^{1/f_E} |\dot{\mathbf{u}}_{\text{pr}}|^2 dt$$
$$= \frac{v\mu_0}{\gamma M_S kT}\oint_E \dot{\mathbf{M}}_{\text{pr}} \cdot d\mathbf{M} \tag{27}$$

and (see Appendix B)

$$V_E = -\frac{v\mu_0 M_S}{f_E \gamma kT}\overline{|\dot{\mathbf{u}}_{\text{pr}}|^2}\frac{\partial P}{\partial E} = \frac{J}{f_E}\overline{\mathbf{e}_P \cdot [\mathbf{u} \times \dot{\mathbf{u}}_{\text{pr}}]} = \frac{J}{f_E}\int_0^1 \mathbf{e}_P \cdot [\mathbf{u} \times \dot{\mathbf{u}}_{\text{pr}}] d\phi$$
$$= J\int_0^{1/f_E} \mathbf{e}_P \cdot [\mathbf{u} \times \dot{\mathbf{u}}_{\text{pr}}] dt = J\oint_E \mathbf{e}_P \cdot [\mathbf{u} \times d\mathbf{u}_{\text{pr}}]. \tag{28}$$

This is the VLD energy-controlled diffusion equation for classical spins including STT. Thus, we have a simple physical method of deriving equation (26), mirroring the Kramers VLD calculation for point particles [1].

**4. Escape from a potential well**

The mean first passage time $\tau_{VLD}$ (i.e., the time to reach the separatrix for the first time from a minimum within a potential well provided that all spins there are absorbed, which is the boundary condition that $W$ vanishes at the critical or separatrix energy) is then, by the quasi-stationary solution of Eq. (26) as given in Ref. [12],

$$\tau_{VLD} = \frac{1}{\alpha}\int_{E_A}^{E_C}\frac{1}{S_E} e^{E - \frac{1}{\alpha}\int_{E_A}^{E}\frac{V_{E''}}{S_{E''}}dE''}\int_{E_A}^{E}\frac{1}{f_{E'}} e^{-E' + \frac{1}{\alpha}\int_{E_A}^{E'}\frac{V_{E''}}{S_{E''}}dE''} dE'dE. \tag{29}$$

Here $E_C$ is the *critical energy* at which a spin can escape a well (by virtue of a thermal fluctuation). The *escape time* is then $2\tau_{VLD}$ and the escape rate $\Gamma$ from a single well is then $\frac{1}{2\tau_{VLD}}$ where $S_E$, $V_E$, and $f_E$ are always determined via the *deterministic* Larmor equation, that is from the unperturbed (lossless) solution [13,16]. In the high barrier limit, which is the only case of interest as detailed in Ref. [16], we have

$$\tau_{VLD}^{as} \sim \frac{e^{\Delta E}}{\alpha f_A S_{E_C}} \qquad (30)$$

where the effective barrier height is given by

$$\Delta E = E_C - E_A - \frac{1}{\alpha} \int_{E_A}^{E_C} \frac{V_E}{S_E} dE. \qquad (31)$$

Furthermore, for a double-well of the potential, if $A$ and $B$ denote the two minima of the potential, it follows that the overall relaxation time $\tau$ is

$$\tau = \frac{2\tau_{VLD}^A \tau_{VLD}^B}{\tau_{VLD}^A + \tau_{VLD}^B}, \qquad (32)$$

where $\tau_{VLD}^A$ and $\tau_{VLD}^B$ denote the escape times from the wells $A$ and $B$, respectively.

## 5. Conclusions

Equation (26)-(28) agree in all respects with the energy-controlled-diffusion equation derived by Apalkov and Visscher [14] via appropriate manipulation of the magnetization vector, and with that derived by Dunn *et al.* [24] by transforming Brown's Fokker-Planck equation to energy $E$ and phase $\phi$ variables using essentially the method of Stratonovich [17] (as reviewed by us in in the zero STT case in Ref. [13]) and so automatically involving the properties of multiplicative noise. For a biaxial potential $vV/(kT) = \sigma(\delta \cos^2 \vartheta - \sin^2 \vartheta \cos^2 \varphi)$ Taniguchi et al [22] used these equations to derive the mean first passage time. This was then extended by Byrne *et al.* [16] to include an external magnetic field of arbitrary strength. Furthermore, the VLD calculations were extended to include the Kramers turnover region in Kalmykov *et al*. [15].

Thus, the original VLD approach of Kramers (1940) based on calculating directly the averages over slightly perturbed orbits, when applied to magnetic nanoparticles, yields both a simpler and more easily visualized solution of the VLD problem than those hitherto existing.

## Acknowledgements

D. J. Byrne acknowledges Science Foundation Ireland for financial support. This publication has emanated from research conducted with the financial support of Science Foundation Ireland under Grant number 17/IFB/5420.

**Appendix A: Changing the variables in St(*W*)**

We now consider the right-hand side of the Fokker-Planck equation (13) for the surface probability function $W(\vartheta,\varphi,t)$. Here we also use the transformation to energy and phase variables, viz., $\{\vartheta,\varphi\} \to \{E,\phi\}$ separating St(*W*) in two parts $St(W) = St(W)|_E + St(W)|_\phi$, where $St(W)|_E$ and $St(W)|_\phi$ involves derivatives with respect to *E* and and $\phi$, respectively. First consider $St(W)|_E$ involving the derivative with respect to energy via $\frac{\partial}{\partial \vartheta} = \frac{\partial E}{\partial \vartheta}\frac{\partial}{\partial E}$ and $\frac{\partial}{\partial \varphi} = \frac{\partial E}{\partial \varphi}\frac{\partial}{\partial E}$ so that $St(W)|_E$ becomes

$$St(W)|_E = \alpha \frac{\gamma kT}{v\mu_0 M_S} \frac{1}{\sin\vartheta}\left\{\frac{\partial E}{\partial \vartheta}\frac{\partial}{\partial E}\left(\sin\vartheta \frac{\partial E}{\partial \vartheta}\left(\frac{\partial W}{\partial E} + W + \alpha^{-1}\frac{\partial P}{\partial E}W\right)\right) \right.$$
$$\left. + \frac{1}{\sin\vartheta}\frac{\partial E}{\partial \varphi}\frac{\partial}{\partial E}\left[\frac{\partial E}{\partial \varphi}\left(\frac{\partial W}{\partial E} + W + \alpha^{-1}\frac{\partial P}{\partial E}W\right)\right]\right\}. \quad (A1)$$

Now considering the gyromagnetic equation (1), written explicitly in the spherical polar coordinate system as

$$\dot{\mathbf{u}}_{pr} = \frac{\gamma kT}{v\mu_0 M_S}\left(-\frac{1}{\sin\vartheta}\frac{\partial E}{\partial \varphi}\mathbf{e}_\vartheta + \frac{\partial E}{\partial \vartheta}\mathbf{e}_\varphi\right), \quad (A2)$$

where $\mathbf{e}_\vartheta$ and $\mathbf{e}_\varphi$ are the basis vectors, using Eq. (18) and noticing that $|\dot{\mathbf{u}}_{pr}| = |\dot{\phi}\sin\vartheta|$,

$$\left(\dot{\mathbf{u}}_{pr}\cdot\frac{\partial \dot{\mathbf{u}}_{pr}}{\partial E}\right) = \frac{1}{2}\frac{\partial}{\partial E}\left(\dot{\mathbf{u}}_{pr}\cdot\dot{\mathbf{u}}_{pr}\right) = \frac{1}{2}\frac{\partial}{\partial E}\left(|\dot{\mathbf{u}}_{pr}|^2\right) = |\dot{\mathbf{u}}_{pr}|\frac{\partial |\dot{\mathbf{u}}_{pr}|}{\partial E}, \quad (A3)$$

$$\dot{\mathbf{u}}_{pr}\cdot\frac{\partial}{\partial E}\left(\dot{\mathbf{u}}_{pr}F\right) = |\dot{\mathbf{u}}_{pr}|^2\frac{\partial F}{\partial E} + F\left(\dot{\mathbf{u}}_{pr}\cdot\frac{\partial \dot{\mathbf{u}}_{pr}}{\partial E}\right), \quad (A4)$$

where

$$F = \frac{|\dot{\mathbf{u}}_{pr}|}{f_E}\left(\frac{\partial f_E W}{\partial E} + f_E W + \alpha^{-1} f_E \frac{\partial P}{\partial E} W\right), \tag{A5}$$

Eq. (A1) becomes in the new variables $\{E, \phi\}$

$$\text{St}(f_E W)\big|_E = \alpha \frac{v\mu_0 M_S}{\gamma kT} f_E \frac{\partial}{\partial E}\left(\frac{|\dot{\mathbf{u}}_{pr}|^2}{f_E}\left(\frac{\partial}{\partial E}(f_E W) + f_E W + \alpha^{-1} f_E \frac{\partial P}{\partial E} W\right)\right). \tag{A6}$$

The change of variables in the term $\text{St}(W)\big|_\phi$ can be accomplished in like manner. However, on averaging $\text{St}(W)\big|_\phi$ over the *fast* phase variable $\phi$, we may *drop all terms in $\overline{\text{St}(W)\big|_\phi}$ containing the phase derivative $\partial_\phi$ because any averaged function* $\bar{F}(E,\phi)$ will not depend on the phase $\phi$, that is $\bar{F}(E,\phi) = \bar{F}(E)$ yielding $\partial_\phi \bar{F}(E) = 0$. Thus $\overline{\text{St}(W)\big|_\phi} = 0$.

**Appendix B: Work done by the STT**

We demonstrate that the STT current induced term of the energy-controlled-diffusion equation can be expressed in the conventional form as the work $V_E$ done by the STT, Eq. (28),

$$\frac{v\mu_0 M_S}{\gamma kT}\frac{\partial}{\partial E}\left(\overline{|\dot{\mathbf{u}}_{pr}|^2 \frac{\partial P}{\partial E}} W\right) = -\frac{\partial}{\partial E}(V_E f_E W). \tag{B1}$$

First we rewrite the averaged term using Eqs. (3) and (4) as

$$\overline{|\dot{\mathbf{u}}_{pr}|^2 \frac{\partial P}{\partial E}} = \left(\frac{\gamma kT}{v\mu_0 M_S}\right)^2 \overline{\frac{\partial E}{\partial \mathbf{u}} \cdot \frac{\partial E}{\partial \mathbf{u}} \frac{\partial P}{\partial E}} = \left(\frac{\gamma kT}{v\mu_0 M_S}\right)^2 \overline{\frac{\partial E}{\partial \mathbf{u}} \cdot \frac{\partial P}{\partial \mathbf{u}}}. \tag{B2}$$

Using the gyromagnetic equation (1)

$$\mathbf{M} \times \dot{\mathbf{M}}_{pr} = \gamma \mathbf{M} \times (\mathbf{H} \times \mathbf{M}) = \gamma M_S^2 \mathbf{H} = -\frac{\gamma kT M_S}{\mu_0 v}\frac{\partial E}{\partial \mathbf{u}} \tag{B3}$$

and our simplified form of the STT induced nonconservative potential, Eqs. (8) and (12),

$$\frac{\partial P}{\partial \mathbf{u}} = \frac{v}{kT}\frac{\partial \Phi}{\partial \mathbf{u}} = J\mathbf{e}_P, \tag{B4}$$

Eq. (B2) can be written as

$$\overline{|\dot{\mathbf{u}}_{pr}|^2 \frac{\partial P}{\partial E}} = -J\frac{\gamma kT}{v\mu_0 M_S}\overline{\mathbf{e}_P \cdot [\mathbf{u} \times \dot{\mathbf{u}}_{pr}]}, \tag{B5}$$

thus justifying Eqs. (B1) and (28).